\documentclass[11pt]{article}
\usepackage{moriond,epsfig}
\usepackage{amsmath}
\usepackage{graphicx}
\usepackage{amsbsy}
\usepackage{cite}
\usepackage{amsfonts}
\usepackage{amssymb}

\def\ra{\rightarrow}

\def\be{\begin{equation}}
\def\ee{\end{equation}}
\def\bea{\begin{eqnarray}}
\def\eea{\end{eqnarray}}

\begin{document}
\vspace*{4cm}
\title{$K^+ \ra \pi^+ \pi^0 \gamma$ in the Standard Model and Beyond}

\author{ P. MERTENS }

\address{Centre for Cosmology, Particle Physics and Phenomenology (CP3)\\
Universit\'e catholique de Louvain, 
Chemin du Cyclotron, 2\\
B-1348 Louvain-la-Neuve, Belgium}

\maketitle\abstracts{In this note we show how improved theoretical analysis combined with recent experimental data coming from NA48/2 concerning $K^+ \to \pi^+ \pi^0 \gamma$ decay shed light on the dynamics of the $s \rightarrow d \gamma$ transition. Consequences on NP analysis are also presented.}

\section{Introduction}
In the search for New Physics (NP) the $s\rightarrow d\gamma$ process is complementary to $b\rightarrow s\gamma$ and $\mu\rightarrow e\gamma$, as the relative strength of these transitions is a powerful tool to investigate the NP dynamics. However, since $s\rightarrow d\gamma$ takes place deep within the non-perturbative regime of QCD we have to control hadronic effects and find observables sensitive to the short-distance dynamics, and thereby to possible NP contributions. The purpose of this note is to show how this can be achieved using the $K^+ \to \pi^+ \pi^0 \gamma$ observable \cite{CDI99}.

In section 2, the anatomy of the $s\rightarrow d\gamma$ process in the Standard Model (SM) is shortly detailed. In section 3, we analyse the $K^+ \ra \pi^+ \pi^0 \gamma$ decay in the SM whereas section 4 is devoted to show how, in the MSSM, rare and $K^+ \ra \pi^+ \pi^0 \gamma$ decays, as well as $\operatorname{Re}(\varepsilon_K^{\prime}/\varepsilon_K)$ can be exploited to constrain NP.

\section{The $s \ra d \gamma$ anatomy}
In the SM, the flavour changing electromagnetic process $s \ra d \gamma$ is a loop effects which at low energy scale is described by the effective $\Delta S =1$ Hamiltonian~\cite{BuchallaBL96}%
\begin{equation}
\mathcal{H}_{eff}(\mu \approx 1\text{ GeV})=\sum_{i=1}^{10}C_{i}\left(  \mu\right)  Q_{i}\left(  \mu\right)  +C_{\gamma^{\ast}}^{\pm}Q_{\gamma^{\ast}}^{\pm}+C_{\gamma}^{\pm}Q_{\gamma}^{\pm}+h.c.\;, \label{OPE}
\end{equation}
where the $Q_{i}$ are effective four-quarks operators whereas the quark-bilinear electric $Q_{\gamma^{\ast}}^{\pm}$ and magnetic $Q_{\gamma}^{\pm}$ operators are respectively given by\footnote{By definition : $2\sigma^{\mu\nu}=i[\gamma^{\mu},\gamma^{\nu}]$.}
$
Q_{\gamma^{\ast}}^{\pm}=(\bar{s}_{L}\gamma^{\nu}d_{L}\pm\bar{s}_{R}\gamma^{\nu}d_{R})\,\partial^{\mu}F_{\mu\nu}$ and $Q_{\gamma}^{\pm}=(\bar{s}_{L}\sigma^{\mu\nu}d_{R}\pm\bar{s}_{R}\sigma^{\mu\nu}d_{L})\,F_{\mu\nu}$.
In the non perturbative regime of QCD this Hamiltonian is hadronized into an effective weak Lagrangian that shares the chiral properties of the operators contained in $\mathcal{H}_{eff}$. The chiral structures of $Q_{i}$ and $Q_{\gamma^{\ast}}^{\pm}$ allow the usual $\mathcal{O}(p^2)$ weak Lagrangian $\mathcal{L}_{W}=G_8 O_{8} +G_{27}O_{27} +G_{ew}O_{ew}$ (detailed in \cite{us}) whereas the chirality flipping $Q_{\gamma}^{\pm}$ operators induce more involved $\mathcal{O}(p^4)$ local interactions (detailed in \cite{CDI99,us}). The non-trivial dynamics corresponding to the low-energy tails of the photon penguins arise at $\mathcal{O}(p^{4})$ (the $\mathcal{O}(p^{2})$ dynamics being completely predicted by Low's theorem~\cite{Low}) where they are represented in terms of non-local meson loops, as well as additional $\mathcal{O}(p^{4})$ local effective interactions, in particular the $\Delta I=1/2$ enhanced $N_{14},...,N_{18}$ octet counterterms~\cite{CT1,CT2}.

\section{$K^+ \ra \pi^+ \pi^0 \gamma$ in the SM}
For the $K^{+}\rightarrow\pi^{+}\pi^{0}\gamma$ decay,
the standard phase-space variables are chosen as the $\pi^{+}$ kinetic energy
$T_{c}^{\ast}$ and $W^{2}\equiv(q_\gamma \cdot P_K)(q_\gamma\cdot P_{\pi^+})/m_{\pi^{+}}^{2}%
m_{K}^{2}$~\cite{Christ67}. Indeed, pulling out the dominant bremsstrahlung contribution, the differential rate can be written%
\begin{equation}
\frac{\partial^{2}\Gamma}{\partial T_{c}^{\ast}\partial W^{2}}=\frac{\partial
^{2}\Gamma_{IB}}{\partial T_{c}^{\ast}\partial W^{2}}\left(  1-2\frac{m_{\pi
^{+}}^{2}}{m_{K}}\operatorname{Re}\left(  \frac{E_{DE}}{eA_{IB}}\right)
W^{2}+\frac{m_{\pi^{+}}^{4}}{m_{K}^{2}}\left(  \left|  \frac{E_{DE}}{eA_{IB}%
}\right|  ^{2}+\left|  \frac{M_{DE}}{eA_{IB}}\right|  ^{2}\right)
W^{4}\right)  \;. \label{DiffRate}%
\end{equation}
In this expression both electric $E_{DE}$ and magnetic $M_{DE}$ direct emission amplitudes are functions of $W^{2}$ and $T_{c}^{\ast}$ and appear at $\mathcal{O}(p^4)$. To a very good  approximation we can identify these direct emission amplitudes with their first multipole for which the $\pi^+ \pi^0$ state is in a P wave.
The main interest of $K^{+}\rightarrow\pi^{+}\pi^{0}\gamma$ is that its bremsstrahlung component $A_{IB}=A(K^+ \ra \pi^+ \pi^0)$ is pure $\Delta I=3/2$ hence suppressed, making the direct
emission amplitudes easier to access. The magnetic amplitude $M_{DE}$ is dominated by the
QED anomaly and will not concern us here.
\subsection{Differential rate}

Given its smallness, we can assume the absence of CP-violation when discussing
this observable. Experimentally, the electric and magnetic amplitudes (taken
as constant) have been fitted in the range $T_{c}^{\ast}\leq80$ MeV and
$0.2<W<0.9$ by NA48/2~\cite{ExpKppg}.
For the electric amplitude, using their parametrization, we obtain at
$\mathcal{O}(p^{4})$ :%
\begin{equation}
X_{E}=\frac{-\operatorname{Re}\left(  E_{DE}/eA_{IB}\right)  }{m_{K}%
^{3}\cos(\delta_{1}^{1}-\delta_{0}^{2})}=\frac{3G_{8}/G_{27}}{40\pi^{2}F_{\pi}^{2}m_{K}^{2}}\left[
E^{l}(W^{2},T_{c}^{\ast})-\dfrac{m_{K}^{2}\operatorname{Re}\bar{N}}%
{m_{K}^{2}-m_{\pi}^{2}}\right]\equiv X_E^l - X_E^{CT}  \;, \label{Eloop}%
\end{equation} where $\delta_1^1$ ($\delta_0^2$) is the strong phase of $E_{DE}$ ($A_{IB}$).
The $E^{l}$ represents $O_{8}$ and $O_{27}$ induced loop contributions (loop contributions from $O_{ew}$ are sub-leading) and $\bar{N}$ corresponds to local counterterms and $Q_\gamma^-$ contributions. Naively we would expect the $O_{27}$ contributions to be sub-dominant, however, they are dynamically enhanced by $\pi \pi$ loops.
Since experimentally, no slope were included in $X_E$, we average $E^{l}$ over
the experimental range and find $X_{E}^{l}=-17.6\;\mathrm{GeV}^{-4}$.
Knowing $X_{E}^{l}$ and using the experimental measurement of $X_{E}=\left(  -24\pm4\pm4\right)
\;\text{GeV}^{-4}$ we can extract the local contributions 
\begin{equation}
X_E^{CT}/X_E^{l}=0.37\pm 0.32~~\ra ~~ \operatorname{Re}\bar{N}=0.095\pm0.083\;. \label{CT}%
\end{equation}
To our knowledge it is the first time that $K^+ \ra \pi^+ \pi^0 \gamma$ counterterms contributions are extracted from experiment. The value we found is much smaller than the $\mathcal{O}(1)$ expected for the $N_{i}$ on
dimensional grounds or from factorization~\cite{EPRKppg}. Note that the required amount of counterterm contribution would have been bigger if $O_{27}$ loops were neglected since then $X_{E}^{l}=-10.2\;\mathrm{GeV}^{-4}$. This result is important since it implies that the counterterms combination $\bar{N}$, which appears in other radiative $K$ decays, is now under control and further reliable theoretical investigations can be carried on, in particular concerning the CP violating observables.  
\subsection{Direct CP-violating asymmetry}
Since the bremsstrahlung and direct emission amplitudes interfere and carry different strong and weak phases, a non vanishing CP violating asymmetry can be generated. The asymmetry measures direct CP violation since $K^\pm$ do not mix. Besides and because 
the long-distance bremsstrahlung amplitude dominates the branching, this CP asymmetry is the simplest window on short-distance physics and a fortiori on possible NP effects. CP-violation in $K^{+}\rightarrow\pi^{+}\pi^{0}\gamma$ is quantified by the
parameter $\varepsilon_{+0\gamma}^{\prime}$, defined from%
\begin{equation}
\operatorname{Re}\left(  \frac{E_{DE}}{eA_{IB}}\right)  \left(  K^{\pm
}\rightarrow\pi^{\pm}\pi^{0}\gamma\right)  \approx\frac{\operatorname{Re}%
E_{DE}}{e\operatorname{Re}A_{IB}}\left[  \cos(\delta_{1}^1-\delta_{0}^{2}%
)\mp\sin(\delta_{1}^1-\delta_{0}^{2})\varepsilon_{+0\gamma}^{\prime}\right]
\;, \label{Asym1}%
\end{equation}
as $
\varepsilon_{+0\gamma}^{\prime}\equiv \mathrm{Arg} E_{DE}-\mathrm{Arg}A_{IB}$ (see \cite{DAmbrosioI96}). Both $Q_\gamma^-$ and $Q_i$ (through loops and counterterms) contribute to this parameter and we find 
\begin{equation}
\varepsilon_{+0\gamma}^{\prime}(Q_i)=-0.55(25)\frac{\sqrt{2}|\varepsilon^\prime_K|}{\omega}~~\mathrm{and}~~\varepsilon_{+0\gamma}^{\prime}(Q_{\gamma}^{-})=+2.8(7)\frac{ \operatorname{Im}C_{\gamma}^{-}}{G_{F}m_{K}}\;,\label{dgf}
\end{equation}
respectively\footnote{ Numerically, in the SM, the Wilson
coefficient of the magnetic operator in $b\rightarrow s\gamma$ can be used for
$\operatorname{Im}C_{\gamma}^{\pm}$, since the CKM elements for the $u$, $c$,
and $t$ contributions scale similarly and we find $
\operatorname{Im}C_{\gamma}^{\pm}(2\;\text{GeV})_{\text{SM}}/{G_{F}%
m_{K}}=\mp0.31(8)\times
\operatorname{Im}\lambda_{t}
$.}. Sadly, these contributions interfere destructively implying that  $\varepsilon_{+0\gamma}^{\prime
}|_{\text{SM}}=0.5(5)\times10^{-4}$. This large uncertainty is driven by a large uncertainty on counterterms and on estimated $\mathcal{O}(p^6)$ effects. However, contrary to what happens in $\varepsilon^\prime_K$, $\varepsilon_{+0\gamma}^{\prime
}$ is rather insensitive to isospin breaking effects, conservatively taken into account in (\ref{dgf}). Expressing $\varepsilon_{+0\gamma}^{\prime}(Q_i)$ in term of the experimental $\varepsilon^\prime_K$ allows us to keep possible NP effects in $Q_i$ under control. As a consequence, the only way for NP to affect $\varepsilon_{+0\gamma}^{\prime}$ is via its $\operatorname{Im}C_{\gamma}^{-}$ component.
The current bound obtained by NA48/2~\cite{ExpKppg} is rather weak and allows very large NP effects in $\varepsilon_{+0\gamma}^{\prime}$ :
\begin{equation}
{\operatorname{Im}C_{\gamma}^{-}|_{NP}}/{G_{F}m_{K}}=-0.08\pm0.13\;. \label{ep0gExp}%
\end{equation}

\section{$K^+ \ra \pi^+ \pi^0 \gamma$ beyond the SM}
Once combined with other short-distance sensitive observables, any experimental improved measurement of $\varepsilon_{+0\gamma}^{\prime}$ will be greatly rewarding. The main problem when probing NP is the issue of disentangling correlations between various NP sources in a fully model-independent way. In \cite{us}, we analysed broad classes of NP scenarios defined as model-independently as possible and identified corresponding strategies to constrain and disentangle NP sources using experimental informations on $K_L \ra \pi \ell^+ \ell^-$, $K \ra \pi \nu \bar{\nu}$ decays and $\mathrm{Re}(\varepsilon^\prime_K / \varepsilon_K)$. Doing so we highlighted the complementary informations that could be obtained from radiative decays. 

\begin{figure}[t]
\centering
\begin{tabular}
[b]{cc}%
$a.\text{\raisebox{-3.0cm}{\includegraphics[width=7.0cm]{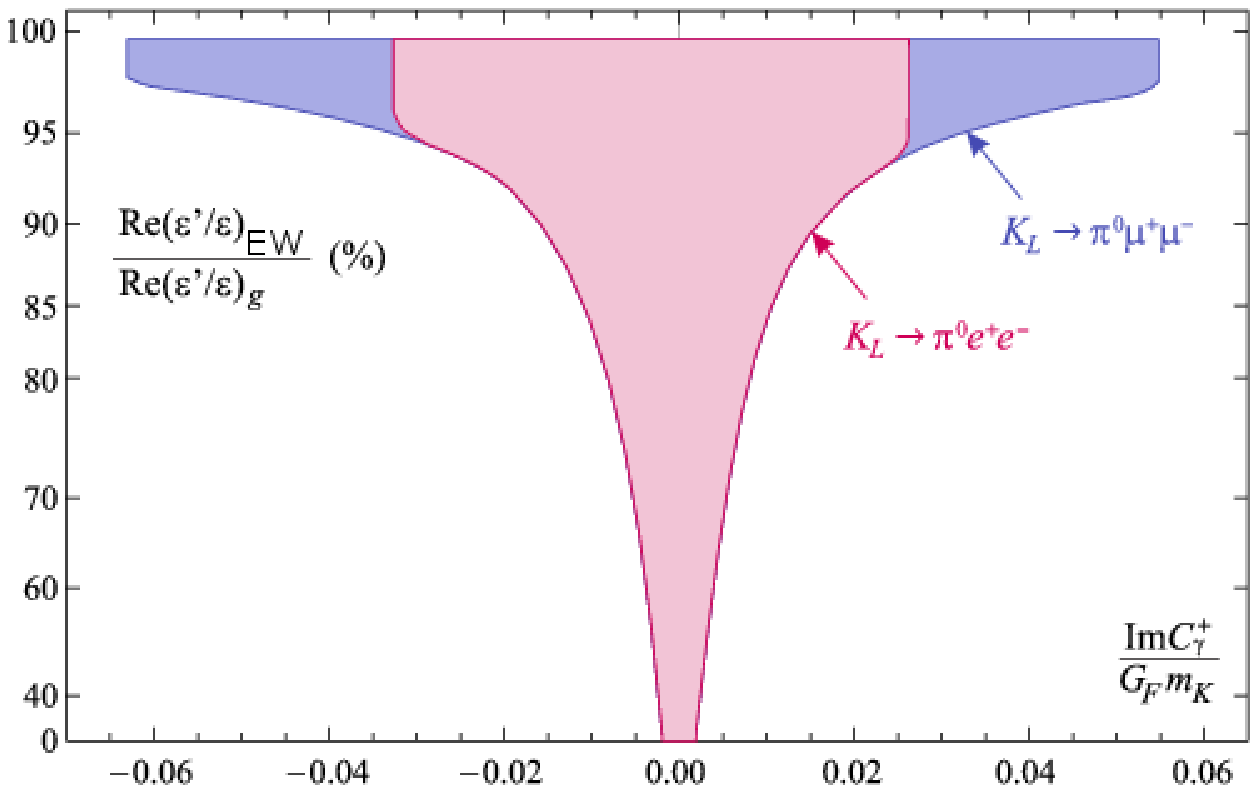}}  }$&
$b.\text{\raisebox{-3.0cm}{\includegraphics[width=7.0cm]{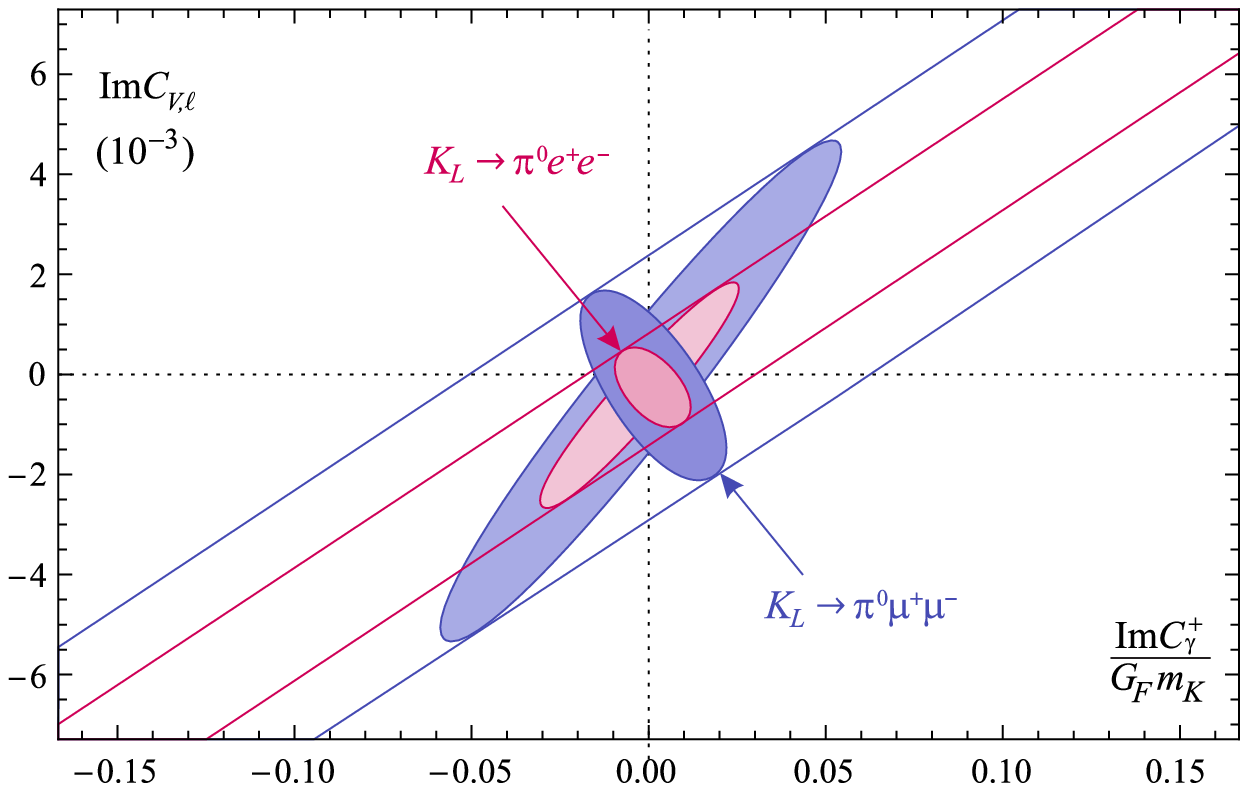}}  }$
\end{tabular}
\caption{Loop-level FCNC scenario, with all the electroweak operators as well
as $Q_{\gamma,g}^{\pm}$ simultaneously turned on, but imposing
$\operatorname{Im}C_{\gamma}^{+}=\pm1.5\operatorname{Im}C_{g}^{-}$ and  $|\operatorname{Re}(\varepsilon^{\prime
}_K/\varepsilon_K)^{\text{NP}}|<2\operatorname{Re}(\varepsilon^{\prime
}_K/\varepsilon_K)^{\exp}$. ($a$) The $\operatorname{Im}C_{\gamma}^{+}$ range as a
function of the fine-tuning between $\operatorname{Re}(\varepsilon^{\prime
}_K/\varepsilon_K)_{EW}$ and $\operatorname{Re}(\varepsilon^{\prime}%
/\varepsilon)_{g}$. ($c$) The corresponding contours in the $\operatorname{Im}%
C_{V,\ell}-\operatorname{Im}C_{\gamma}^{+}$ plane. In ($b$), the
lighter (darker) colors denote destructive (constructive) interference between
NP $\gamma^*$-penguin and $Q_{\gamma}^{+}$ in $K_{L}\rightarrow\pi^{0}\ell^{+}\ell^{-}$.}%
\label{Fig2}%
\end{figure}

In the MSSM \cite{BurasS98,BurasCIRS99,MST06,ColangeloI98,IsidoriMPST06,BurasEJR05}, NP can affect all the operators in (\ref{OPE}) as well as gluon-penguin (denoted by $Q_g^\pm$) and semi-leptonic operators, in particular $Q_{V,l}=\bar{s}\gamma_\mu d \otimes \bar{\ell} \gamma^\mu \ell$. In this particular model the irreducible correlations are two fold. First $Q_{\gamma}^+$ and $Q_{V,l}$ ($\ni Q_{\gamma^\ast}^+$) always interfere in $K_L \ra \pi \ell^+ \ell^-$ in and beyond the SM and second, $\mathrm{Re}(\varepsilon^\prime_K / \varepsilon_K)$ receives NP contributions from many different sources. The corresponding bounds are displayed in Figure \ref{Fig2} where we see that a large but not impossible cancellation between NP in gluon-penguin and electroweak operators in $\mathrm{Re}(\varepsilon^\prime_K / \varepsilon_K)$ allows for $\mathrm {Im}C_\gamma^+$ to reach the percent level if we impose $\operatorname{Im}C_{\gamma}^{+}=\pm1.5\operatorname{Im}C_{g}^{-}$. This value will correspond to a saturation of the current $K_L \ra \pi^0 e^+ e^-$ upper bound and since in the MSSM $Q_\gamma^\pm$ and $Q_g^\pm$ mix under renormalization this $\mathrm {Im}C_\gamma^+$ upper bound provides also an lower bound for $\mathrm {Im}C_\gamma^-$. 
From (\ref{dgf}) this implies that NP can push $\varepsilon_{+0\gamma}^{\prime}$ up to roughly two orders of magnitude above its SM prediction. The parameter $\varepsilon_{+0\gamma}^{\prime}$ provides therefore a very good probe for NP $\gamma$-penguin effects and furthermore reveals NP cancellations occurring inside $\mathrm{Re}(\varepsilon^\prime_K / \varepsilon_K)$. 
\section{Conclusion}
We exemplify in $K^+ \ra \pi^+ \pi^0 \gamma$ that the stage is now set theoretically to fully exploit the $s\rightarrow d\gamma$ transition. The SM predictions are under good control, the sensitivity to NP is excellent, and signals in rare and radiative $K$ decays not far from the current experimental sensitivity are possible. Thus, with the advent of the next generation of $K$ physics experiments (NA62 at CERN, K0TO at J-Parc, ORKA at Fermilab and KLOE-II at the
LNF), the complete set of flavor changing electromagnetic processes, $s\rightarrow d\gamma$, $b\rightarrow(s,d)\gamma$, and $\ell\rightarrow\ell^{\prime}\gamma$, could become one of our main windows into the flavor sector of the NP which
will hopefully show up at the LHC.

\section*{Acknowledgments}
I thank the organizers of the \textit{Rencontres de Moriond EW 2012} for the pleasant and stimulating stay and for their financial support. I warmly thank Christopher Smith for the fruitful collaboration at the origin of this work. I'm also grateful to him and Jean-Marc G\'erard for their valuable suggestions and comments about the present note.

\end{document}